\documentclass[%
preprint,
amsmath,amssymb,
aps,
]{revtex4-1}
\usepackage{xcolor}
\usepackage{graphicx}
\usepackage{dcolumn}
\usepackage{bm}
\usepackage{color}

\DeclareMathOperator\erfc{erfc}

\begin{document}
	
	
	\title{Random Surface Statistical Associating Fluid Theory: \\ Adsorption of n-Alkanes on Rough Surface 
	}
	
	\author{Timur Aslyamov}%
	\email{t.aslyamov@gmail.com; taslyamov@slb.com}
	\author{Vera Pletneva}
	\email{vpletneva@slb.com}
	\affiliation{%
		Schlumberger Moscow Research Center; 
		\\13, Pudovkina str., Moscow 119285, Russia}

	\author{Aleksey Khlyupin}
	\email{khlyupin@phystech.edu}
	\affiliation{%
		Moscow Institute of Physics and Technology; 
		\\Institutsky lane 9, Dolgoprudny, Moscow region, 141700, Russia
		\\
		Schlumberger Moscow Research Center; 
		\\ 13, Pudovkina str., Moscow 119285, Russia 
	}%
	\date{\today}
	\begin{abstract}
	Adsorption properties of chain fluids are of interest from both fundamental and industrial points of view. Density Functional Theory (DFT) based models are among the most appropriate techniques allowing to describe surface phenomena. At the same time Statistical Associating Fluid Theory (SAFT) successfully describes bulk PVT properties of chain-fluids. 	In this publication we have developed novel version of SAFT-DFT approach entitled RS-SAFT which is capable to describe adsorption of short hydrocarbons on geometrically rough surface. Major advantage of our theory is application to adsorption on natural roughs surfaces with normal and lateral heterogeneity. For this reason we have proposed workflow where surface of real solid sample is analyzed using theoretical approach developed in our previous work \cite{aslyamov2017density} and experimentally by means of low temperature adsorption isotherm measurements for simple fluids. As result RS-SAFT can predict adsorption properties of chain fluids taking into account geometry of the surface sample under the consideration. In order to test our workflow we have investigated hexane adsorption on carbon black with initially unknown geometry. Theoretical predictions for hexane adsorption at 303K and 293K fit corresponding experimental data well.
\end{abstract}
	
	\pacs{Valid PACS appear here}
	\maketitle
	

\section{Introduction}

Confined fluid thermodynamic properties become crucial when one deals with oil-gas production from unconventional reservoirs, where the fluid often is contained in pores of nanoscopic scale. It is well known that adsorption phenomena play crucial role in capacity characterization of nano-porous materials. For this reason theoretical model which is capable to predict adsorption properties of chain fluid is highly desirable \cite{liu2017adsorption}.  Also industrial needs induce high interest to theory accounting for real adsorbate surface geometry. Indeed, all natural materials are geometrically heterogeneous and nano-roughness has strong influence on surface phenomena \cite{aslyamov2017density}. Thus, in this publication we have proposed theoretical approach describing chain-molecules fluid adsorption on surface with real rough geometry. As result PVT-properties of confined fluid can be described in framework of this theoretical method.  

One of the most useful theoretical approaches to describe thermodynamic properties of real fluid is Statistical Associating Fluid Theory (SAFT) based EOS. Among the variety of known SAFT versions we could emphasize VR-SAFT due to explicit connection with molecular structure and various successful applications to hydrocarbons modeling. Moreover, the model is defined by a set of fluid molecular parameters such as size and characteristic interaction energy. This fact and a wide application motivate the study of PVT-properties for various chain fluids by means of SAFT. In this work we have considered SAFT-VR version for Mie potential, developed by Lafitte et. all \cite{lafitte2013accurate}. This model demonstrates excellent match with experiment for wide range of hydrocarbons \cite{lafitte2013accurate}.

Overwhelming part of publications considering chain molecules in terms of SAFT are dedicated to homogeneous bulk case. However, non-uniform spatial distribution of confined fluids can not be described using bulk EOS only and inhomogeneous extension is needed. In order to take into account impact of spatial boundaries on fluids PVT properties Density Functional Theory (DFT) is often used. Indeed, in literature one can find a variety of surface phenomena which was described in terms of DFT. Despite the fact, that description of simple fluids using DFT has a long history and now is a standard practice, the extension to the case of SAFT-EOS is an actual problem \cite{liu2017adsorption, martinez2017vapour, schindler2013adsorption}. Rigorous way to account for chain structure of molecules in DFT approach is developed in work \cite{yu2002density}. However this model does not consider fluid-fluid interaction which is necessary for accurate description of thermodynamic properties even in bulk case.  Molecular interaction can be described in terms of VR-SAFT \cite{gil1997statistical} where central role plays radial distribution function of hard sphere fluid. This bulk model was used as basis in order to develop inhomogeneous SAFT version describing interfacial behavior of chain molecules. In work \cite{schindler2013adsorption} authors considered adsorption phenomenon using one of the most popular version of DFT and SAFT-VR. Results of this work demonstrate qualitative agreement with computer simulations, however authors did not consider adsorption isotherms and corresponding experimental data. Later DFT-SAFT approach was verified in comparison with experimental data in the case of monomer fluids \cite{malheiro2014density}. In these publications \cite{schindler2013adsorption,malheiro2014density} Radial Distribution Function (RDF) was calculated from result \cite{chang1994real} where analytical expression was obtained, however this form of RDF does not provide correct expression for density derivative which is needed in DFT calculations.  Alternative way was developed in work \cite{franco2017statistical}  where mean-value theorem was applied in order to calculate confinement term in Helmholtz free energy. 
This approach contains an assumption of a square-well potential for the fluid-solid interactions. This fact is serious restriction for further comparison with experiments for natural non-ideal materials, since this simplified potential is unable to describe geometrical roughness. 

In a way similar to the approaches discussed above, we have used VR-SAFT \cite{lafitte2013accurate} as the basis. In another words, our version of SAFT-DFT entitled RS-SAFT (RS means Random Surface) contains VR-SAFT \cite{lafitte2013accurate} as homogeneous bulk limit (far enough from the geometrical boundaries). At the same time RS-SAFT has two major advantages. Firstly, we have developed novel analytical form for fluid-fluid interaction contribution. This expression matches with very popular perturbative expansion from work \cite{lafitte2013accurate} in bulk and can be extended for inhomogeneous fluid. Also this explicit form is derived in terms of Lambert special functions that allows to obtain analytical expression for density derivative. Secondly, we have used our previous results \cite{aslyamov2017density} to account for real surface geometry at nanoscale. Thus, RS-SAFT describes adsorption properties of fluid with accurate bulk-EOS taking into account surface nano-roughness both in normal and lateral directions.    

For wide applications it is necessary to link theoretical predictions with characteristics of real solid samples. Low temperature adsorption analysis is widely used for characterization of pore and surface structures \cite{landers2013density}. Theoretical model can be a part of workflow containing analysis of simple gas adsorption at low temperature and further predictions for more complex fluid adsorption on the same solid sample. For example, in \cite{mitchell2015prediction} authors obtained pore size distribution (PSD) for activated carbon from analysis of nitrogen adsorption at 77K and then calculated n-alkane fluid adsorption taking into account obtained PSD. For adsorption calculations the authors used model from \cite{schindler2013adsorption} which was developed for inhomogeneous fluid near ideal smooth surface.

\begin{figure}
	\includegraphics[width=16cm]{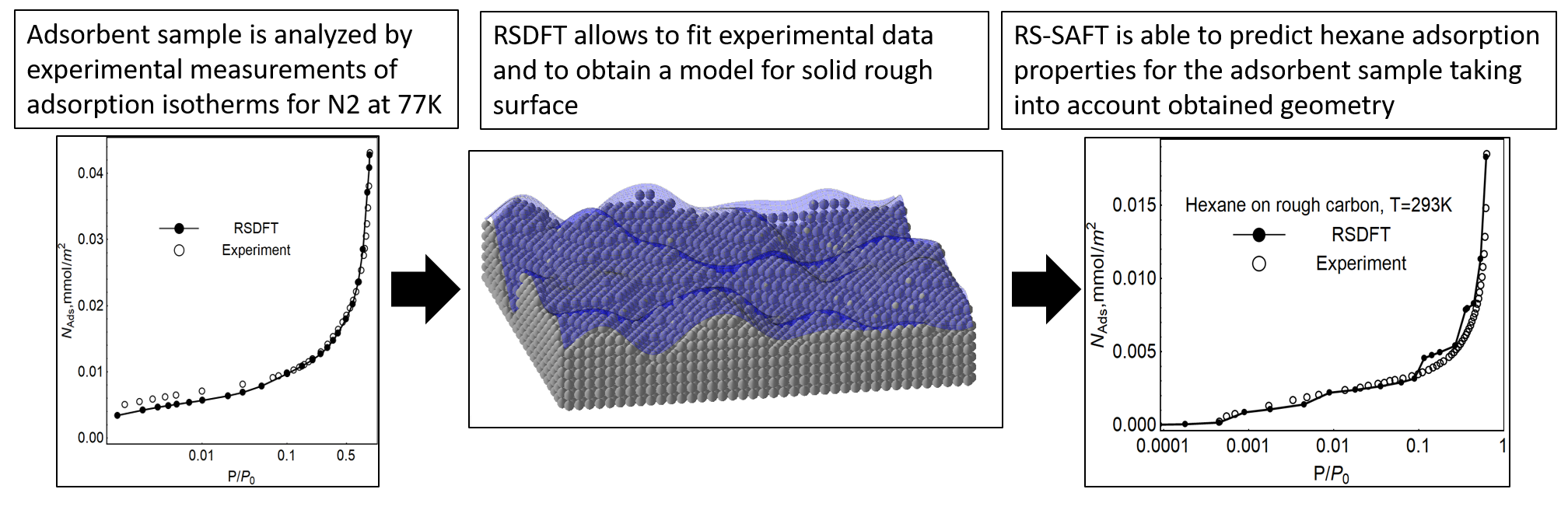}
	\caption{\label{Fig:Workflow} Three major blocks of our workflow: experimental measurement of simple gas adsorption isotherms; RSDFT application allows to fit the data and to model rough surface; RS-SAFT calculations predict adsorption properties for hexane taking into account the surface geometry.}
\end{figure}  

In the current publication we propose the workflow allowing to predict adsorption properties of n-alkanes on natural surface. The major point of this approach is surface geometry analysis of adsorbent sample. In publication \cite{aslyamov2017density} two of us shown that geometrical model can be obtained from adsorption isotherms measurements. The most appropriate way for this procedure is low temperature simple gas adsorption. Thus, experimentally measured adsorption isotherm for simple fluid on adsorbent sample is used as input data for our workflow. In the framework of RSDFT it is possible to fit experimental data and to construct adsorbent surface model corresponding to geometrical roughness of the solid sample. Finally, presented in the current manuscript theory RS-SAFT is capable to describe n-alkanes adsorption properties taking into account obtained surface characteristics. Thus, RS-SAFT workflow allows to predict surface phenomena for chain fluids on natural material using routine experimental data with simple fluids. Example of this workflow can be found in Fig. \ref{Fig:Workflow}. As one can see experimental data for low temperature nitrogen adsorption is used in order to tune RSDFT and to obtain solid surface model. As result RS-SAFT predicts hexane adsorption isotherms on the solid sample using obtained geometry characteristics. 

In order to test our theoretical approach we have investigated hexane adsorption on carbon black with initially unknown surface geometry. In accordance with our workflow we have experimentally measured adsorption isotherm for nitrogen at 77K. We have used RSDFT to obtain geometry roughness characteristics in normal and lateral directions. Finally, we have performed experiments on hexane adsorption on the same solid sample at several temperatures. The comparison demonstrates that RS-SAFT predictions fit experimental results well.

\section{Inhomogeneous version of SAFT}

The general idea is modification of known bulk SAFT Helmholtz free energy using certain version of DFT as was done previously \cite{liu2017adsorption, martinez2017vapour, schindler2013adsorption}. As we discussed above in our study VR-SAFT plays the role of the basic model corresponding to homogeneous limit. More precisely, our bulk fluid corresponds to one-component system of molecules with chain-structure consisting of $M$ spherical segments with diameter $\sigma$. Molecular interaction is defined by monomer fluid-fluid Mie-potential between the segments:
\begin{eqnarray}
U(r)= \frac{\epsilon\lambda_r}{\lambda_r-\lambda_a}\left(\dfrac{\lambda_r}{\lambda_a}\right)^{\frac{\lambda_a}{\lambda_r-\lambda_a}}\left[\left(\dfrac{\sigma}{r}\right)^{\lambda_r}-\left(\dfrac{\sigma}{r}\right)^{\lambda_a}\right]
\end{eqnarray}   
where $r$ is the distance between the segments, $\epsilon$ is the potential depth, $\lambda_a$ and $\lambda_r$ are attractive and repulsive exponents, respectively. In the case of bulk fluid this system was successfully described by VR-SAFT in work \cite{lafitte2013accurate}. In accordance with SAFT formalism Helmholtz free energy can be expressed in terms of ideal, monomer-segment and chain contributions as
\begin{eqnarray}
F=F^{id}+F^{mono}+F^{ch}
\end{eqnarray} 

In formulation of ideal contribution $F^{id}$ we have followed to work \cite{yu2002density} where ideal part contains translational degree of freedom and takes into account chain connectivity via bonding potential. This expression accounts configuration of chain in three dimensional space which is defined by a set of $M$ coordinates $\bold{R=(r_1,..,r_M)}$. Ideal term accounts direct effect of connectivity while chain term $F^{ch}$ corresponds to indirect interactions due to the excluded volume effects. Chain contribution is defined from \cite{lafitte2013accurate} using inhomogeneous density distribution.

In accordance with formulation of VR-SAFT \cite{lafitte2013accurate} we have used thermodynamic perturbation theory for monomer contribution $F^{mono}$:
$$ 
F^{mono}=\sum_{n=0}^{\infty}\beta^n F_n,
$$
where $F_0$ is reference system, $\beta F_1$ and $\beta^2 F_2$ are the first two perturbations terms. It is  possible to map the reference fluid to hard spheres system with another effective diameter \cite{barker1967perturbation2}:
$$
d=\int_{0}^{\sigma}\left(1-e^{-\beta U(r)}\right)dr<\sigma.
$$
Thus, reference system corresponds to gas of hard spheres with new diameter $d$. For inhomogeneous hard sphere contribution $F_0$ we have used result of Fundamental Measure Theory \cite{roth2010fundamental} which equals to corresponding term from \cite{lafitte2013accurate} at bulk limit.

In order to calculate perturbation terms authors of \cite{lafitte2013accurate} used Barker-Henderson theory, where the first two perturbation terms are defined using hard sphere radial distribution function RDF $g(r,\rho)$ only. Thus, explicit expressions for density of Helmholtz free energy  $a_1=\beta F^b_1/N$ and $a_2=\beta F^b_2/N$ ($N$ is the number of molecules) have the following form:
\begin{eqnarray}
\label{a1a2}
&a_1=2\pi\rho \int_{\sigma}^{\infty}g(r)U(r)r^2dr \\
&a_2=-\pi\rho K^{HS}(1+\chi)\int_{\sigma}^{\infty}g(r)(U(r))^2r^2dr \nonumber
\end{eqnarray}
The second expression in \eqref{a1a2} is different from classical one \cite{barker1967perturbation1}, because additional multiple $K^{HS}(1+\chi)$ is applied \cite{lafitte2013accurate}, where $\chi$ is correction pre-factor which depends on density of fluid (we have used expression from \cite{lafitte2013accurate}), $K^{HS}=kT(\partial \rho/ \partial P)_T$ is isothermal compressibility of reference system (HS fluid), $\rho$ and $P$ are the number segment density and the pressure of HS fluid, respectively. Parameter $K^{HS}=K^{HS}(\phi)$ is function of dimensionless density $\phi=\frac{4}{3}\pi\rho d^3$ and can be calculated from the Carnahan Starling compressibility \cite{carnahan1969equation}
$$
K^{HS}=\frac{(1-\phi)^4}{1+4\phi+4\phi^2-4\phi^3+\phi^4},
$$

It is possible to extend results for the first two perturbative terms to the case of inhomogeneous fluids, following the ideas proposed by Toxvaerd \cite{toxvaerd1981structure}. At the same time inhomogeneous case contains serious issues related to the absence of explicit representation for inhomogeneous RDF $g(|r_1-r_2|,\rho(r_1),\rho(r_2))$. However a lot of studies note that there is adequate approximations allowing to consider bulk RDF at averaged density $g(|r_1-r_2|,\bar{\rho})$, which has the following expression:

$$
\bar{\phi}=\frac{\phi(\mathbf{r_1})+\phi(\mathbf{r_2})}{2}
$$
Thus the first two perturbation terms at the case of inhomogeneous fluid can be represented as the follows:
\begin{eqnarray}
\label{A1A2}
&F_1=\dfrac{1}{2}\int d\mathbf{r_1}\int d \mathbf{r_2}\rho(\mathbf{r_1})
\rho(\mathbf{r_2})g(\bar{\phi}(\mathbf{r_1},\mathbf{r_2});|\mathbf{r_1-r_2}|)U(|\mathbf{r_1-r_2}|) \\
&F_2=K^{HS}(1+\chi)\dfrac{1}{2}\int d\mathbf{r_1}\int d \mathbf{r_2}\rho(\mathbf{r_1})
\rho(\mathbf{r_2})g(\bar{\phi}(\mathbf{r_1},\mathbf{r_2});|\mathbf{r_1-r_2}|)U(|\mathbf{r_1-r_2}|)^2 \nonumber
\end{eqnarray}

For further work it is important to note the two points: 1) contrary to the bulk case approximations which were performed in \cite{lafitte2013accurate} can not be used here due to abrupt oscillations of $\rho(\mathbf{r})$ near the walls; 2) DFT approach for chain molecules \cite{yu2002density} demands knowledge of functional derivatives of Helmholtz free energy. For these reasons in this work we have used novel form of RDF allowing to obtain the potentials  \eqref{A1A2} and the density derivative as analytical expressions. 

\begin{figure}
	\includegraphics[width=8cm]{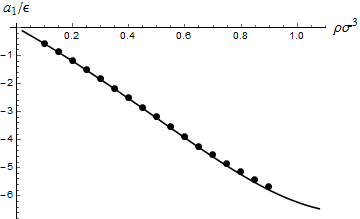}
	\includegraphics[width=8cm]{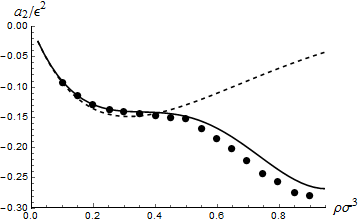}
	\caption{\label{Fig_a1} The density $\rho \sigma^3$ dependence of the first perturbation terms $a_1$ and $a_2$ for the LJ system at temperature $T=\epsilon/k$. Solid curves are analytical results \eqref{a1a2}, dots correspond to Monte Carlo simulations \cite{lafitte2013accurate}.  Dashed curve corresponds to analytical result \eqref{a1a2} without correction prefactor $\chi=0$. }
\end{figure}

RDF is obtained in Percus-Yevick (PY) assumptions and corresponding Verlet Weis (VW) modification have the following form:
\begin{eqnarray}
\label{RDF_Lamberts}
g(r,\phi)=1+\frac{2d}{r}\sum_{n=1}^{K}A_n(\phi)e^{R_n r/d}\cos(I_n r/d+\alpha_n).
\end{eqnarray}
\begin{eqnarray}
\label{RDF_VW}
&g^{VW}(r,\phi)=g(r d/d_M, \phi_M)+\dfrac{1}{r}e^{\frac{\alpha(r-d)}{d}}\cos\dfrac{\alpha (r-d)}{d}
\end{eqnarray}
detailed description of density functions $d_M$,$\phi_M$, $\alpha$,$A_n$, $\alpha_n$, $R_n$, $I_n$ and system parameter $K$, where $n=1,...,K$ can be found in Appendix \ref{Appendix}. This result \eqref{RDF_Lamberts} was obtained as direct calculation of famous Wertheim's solution of Ornstein-Zernike  equation in PY approximation. The final expression was modified in accordance with Verlet-Weiss corrections. As one can see from Fig.~\ref{Fig_a1} expression \eqref{RDF_Lamberts} fits simulations results well. Key difference of this RDF form and alternative published versions is that potentials \eqref{A1A2} and their density derivatives have analytical expressions. 

In order to demonstrate appropriateness of our form of RDF for calculations of perturbations terms we started from bulk case. Integrals \eqref{a1a2} can be calculated analytically using RDF \eqref{RDF_Lamberts}, analytical results can be found in Appendix \ref{Appendix}. As one can see from Fig.~\ref{Fig_a1} these perturbation terms at bulk case have good agreement with numerical results from work \cite{lafitte2013accurate}. 

It is standard to describe inhomogeneous fluids by means of grand canonical thermodynamic potential $\Omega[\rho(\vec{r})]$:
\begin{eqnarray}
\label{GCE}
\Omega[\rho_M(\mathbf{R})]=F[\rho_M(\mathbf{R})]+\int d\mathbf{R}\rho_M(\vec{r})(U_{fs}(\mathbf{R})-\mu)
\end{eqnarray}  
where $U_{fs}$ is the external potential, $\mu$ is the chemical potential, $\rho_M(\mathbf{R})$ is chain configuration density (dependence on $\mathbf{R}$ is crucial difference from the standard DFT approach). Our interest is focused on description of fluids near solid wall, for this reason external potential is defined by interaction between solid molecules and fluid segments. Equilibrium density distribution corresponds to the solution of the following equation:  
\begin{eqnarray}
\label{Eq_1}
\dfrac{\delta \Omega}{\delta \rho_M(\mathbf{R})}=0
\end{eqnarray} 
as was demonstrated in work \cite{yu2002density} in the case when terms of $F^{mono}$ and $F^{chain}$ are depended on segment density $\rho$ solution of \eqref{Eq_1} can be written as function of $\mathbf{r}$. In assumption of surface radial symmetry external potential can be represented as function of one spatial coordinate only $U_{fs}(z)$, where $z$ is the distance between the chain-segment and the surface \cite{Khlyupin2017}. 
\begin{eqnarray}
\label{Lagrange_4}
\rho(z)=e^{\beta \mu}\sum_{i=1}^{M}e^{-\beta\lambda_i(z)}G^{i}(z)G^{M+1-i}(z)
\end{eqnarray}
where $\lambda_i(z)=\dfrac{\delta}{\delta\rho}(F^{mono}+F^{HS})+U_{fs}(z)$  and function $G^i(z)$ is determined from recurrence formula:
\begin{eqnarray}
G^{i}(z)=\int dz' e^{-\beta\lambda_i(z')}\frac{\theta(d-|z-z'|)}{2d}G^{i-1}(z')
\end{eqnarray} 
with $G_1=1$ and $i=2,...,M$. For numerical calculations it is more convenient to consider $\eta(z)=\rho(z)/\rho_0$, where $\rho_0$ is segment bulk density which satisfies to equation \eqref{Lagrange_4} at the limit $z\to\infty$. Let us consider \eqref{Lagrange_4} far enough from the walls that $U_{fs}\simeq0$, that is equivalent to $\lambda_i(0)\equiv\lambda_0$, then one can obtain the following equation: 
\begin{eqnarray}
\label{Lagrange_Bulk}
\rho_0=e^{\beta \mu}e^{-\beta\lambda_0}\sum_{i=1}^{M}e^{-(i-1)\beta\lambda_0}e^{-(M-i)\beta\lambda_0}=Me^{\beta\mu-M\beta\lambda_0}
\end{eqnarray}
After substituting of this result into equation \eqref{Lagrange_4} the following equation can be written:
\begin{eqnarray}
\eta(z)=\frac{1}{M}e^{M\beta\lambda_0}\sum_{i=1}^{M}e^{-\beta\lambda_i(z)}G^{i}(z)G^{M+1-i}(z)
\end{eqnarray}
This equation can be solved by method of simple iteration. 
\section{Surface geometry and external potential}
It already was assumed that the system has radial symmetry and density distribution depends on z-coordinate only. In terms of external potential this situation is equivalent to $U_{fs}\equiv U_{fs}(z)$. Such model allows to investigate gas adsorption on surface of amorphous materials.

Usually natural materials are rough at nano-scale and surface geometry significantly influences the properties of confined fluids. One of the most modern model describing real surface geometry is correlated Gaussian random process \cite{herminghaus2012universal}. In the frame of this approach random rough surface can be characterized by two natural parameters:
\begin{itemize}
	\item Parameter $\delta$ corresponds to variance of the fluctuating surface roughness height in the normal direction.
	\item Parameter $\tau$ (the correlation length) determines the characteristic scale along the lateral direction of the correlation function decay. For example, the correlation length of white noise equals zero, which reflects the independence of heights at any two points in the lateral plane.
\end{itemize} 
In work \cite{Khlyupin2017} two of us obtained fluid-solid potential for rough surface using correlated random model. Our approach is based on Free Energy Averaging Technique presented in the work \cite{forte2014effective}. Also advanced theory of Markovian random processes and the first passage time probability problem were applied as part of averaging procedure (see work \cite{Khlyupin2017} for more details). Proposed model may be applied for wide range of correlation functions of the random solid surfaces. As result potential $U_{fs}(z)$ accounts for surface roughness 
via $\delta, \tau$ and depends on z-coordinate only. Thus, after this modification density distribution equation \eqref{Lagrange_4} is still applicable. Also surface geometry restricts configurational space available for fluid molecule. This fact modifies result of integration $\int dx dy$. One can calculate this two-dimensional integral taking into account only permitted domains for fluids at a certain level $z$ as
\begin{eqnarray}
\label{proper_volume}
\int dz \rho(z)\int_A dxdy...=\int dz \rho(z) S(z) ...
\end{eqnarray}
where $A$ is the total area and $S(z)$ is the part of $A$ which is free from solid media at level $z$. One can find this area from the following expression:
\begin{eqnarray}
S(z)=A\left(1-\frac{1}{2}\erfc\frac{z}{\sqrt{2}\delta}\right)
\end{eqnarray}
Adsorption isotherm in the case of rough surface with known parameters $\delta, \tau$ can be calculated using external potential $U_{fs}(z)$ \cite{Khlyupin2017} and expression \eqref{proper_volume}:
\begin{eqnarray}
N_{ads}=\frac{1}{M}\int d\mathbf{r}\rho(z)-\rho_0 A (H-\sigma_{sf})=\frac{1}{M}\int dz S(z) \rho(z)-\rho_0 A (H-\sigma_{sf})
\end{eqnarray} 

\section{Results}
 In this section we have applied the developed theory for consideration of gas adsorption on surface of carbon black using developed theory. Despite carbon black is often used in adsorption investigations its surface geometry is not clear. A lot of publications claim non-ideal structure of surface at nano-scale. However none of these studies defined the roughness based on direct measurements. For this reason surface characterization of the adsorbate sample is serious challenge.  
 
 Our major aim is describe n-alkane adsorption on certain carbon sample. In order to make desired theoretical predictions taking into account real surface geometry we have proposed the workflow containing  the following two steps:
 \begin{itemize}
 	\item Considered solid sample is investigated experimentally by simple fluid (argon, nitrogen) adsorption at low temperature. RSDFT allows to obtain geometrical characteristics of rough surfaces form adsorption isotherms analysis, both hight variance $\delta$ (roughness in normal direction) and correlation length $\tau$ (lateral characteristic).
 	
 	\item Developed theory RS-SAFT can use obtained from RSDFT characteristics of rough surface geometry in further theoretical predictions of complex (chain) fluid adsorption properties. 
 \end{itemize}
As result the theory can provide accurate PVT-properties in the bulk phase for wide range of n-alkanes and also take into account roughness of adsorbate sample. Finally in this section one can compare theoretical predictions with experimental data for n-alkane adsorption.

\subsection{Analysis of surface geometry}

Simple gas adsorption on ungraphitized carbon black was investigated in work \cite{aslyamov2017density} where RSDFT analysis demonstrated significant geometrical roughness of the surface. In accordance with RSDFT approach, information about surface geometry can be obtained from adsorption isotherms analysis. More precisely, smooth form (without steps) of the isotherm is related to geometrically rough structure of surface. Thus, the best fit of experimental data by RSDFT calculations allows to define roughness parameter $\delta$ and lateral characteristic $\tau$. This information is sufficient to construct surface profile corresponding to investigated solid sample. 

\begin{figure}
	\includegraphics[width=16.0cm]{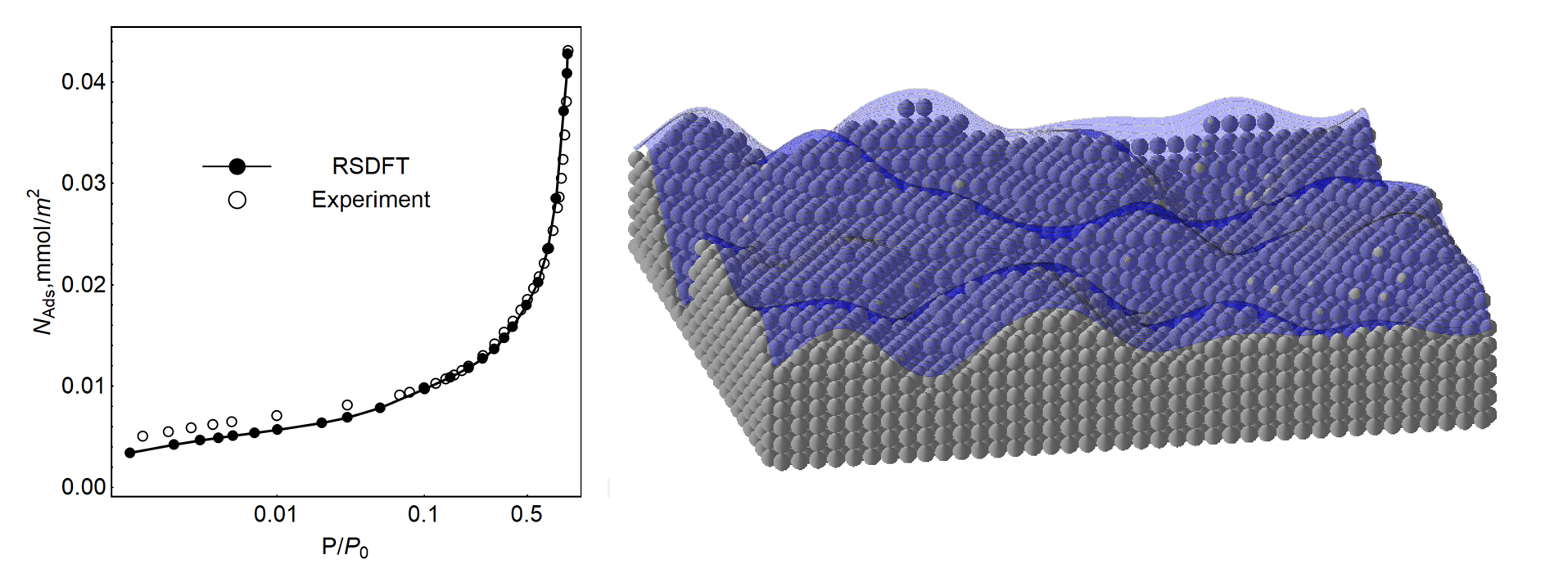}
	\caption{\label{Fig:RSDFT}  Left: Open circles are experimental data for nitrogen adsorption on the sample of carbon black at 77K. Solid line is theoretical results of RSDFT which is tuned by parameters $\delta=0.6$ nm and $\tau=2.1$ nm. Right: example of rough surface corresponding to obtained by RSDFT roughness $\delta=0.6$ nm and $\tau=2.1$ nm.}
\end{figure}

We have performed experimental measurements of surface geometry using ultrahigh-purity nitrogen (99.999\%) as the adsorbate. 
Experimental adsorption isotherms were measured at 77K (Fig.~\ref{Fig:RSDFT}) using an ASAP2020 volumetric adsorption analyzer from Micromeritics, USA and allowed to obtain geometry characteristics of the carbon surface. Carbon black powder provided by Sigma-Aldrich was used as representative carbon material for both theoretical and experimental study.  Prior the isotherm measurements, the sample was subjected to degassed under vacuum at elevated temperatures (473K) overnight (12 h) to remove any adsorbed species. The Brunauer-Emmett-Teller (BET) equation was applied to obtain specific surface area (SSA). For studied carbon black sample it was 1.6 $m^2/g$. The saturation pressure $P_0$ of nitrogen at a liquid nitrogen temperature was determined every two hours during the experiment using a vapor pressure thermometer. 

In order to describe the adsorbate surface we have applied RSDFT approach to obtain adsorption isotherms using the same parameters as in work \cite{aslyamov2017density}. The best fitting of experimental adsorption isotherm by RSDFT corresponds to $\delta=0.6$ nm and $\tau=2.1$ nm. This case is illustrated in Fig.~\ref{Fig:RSDFT}.

\subsection{n-Alkane adsorption on real rough surface}
\begin{figure}[tp]
	\includegraphics[width=12cm]{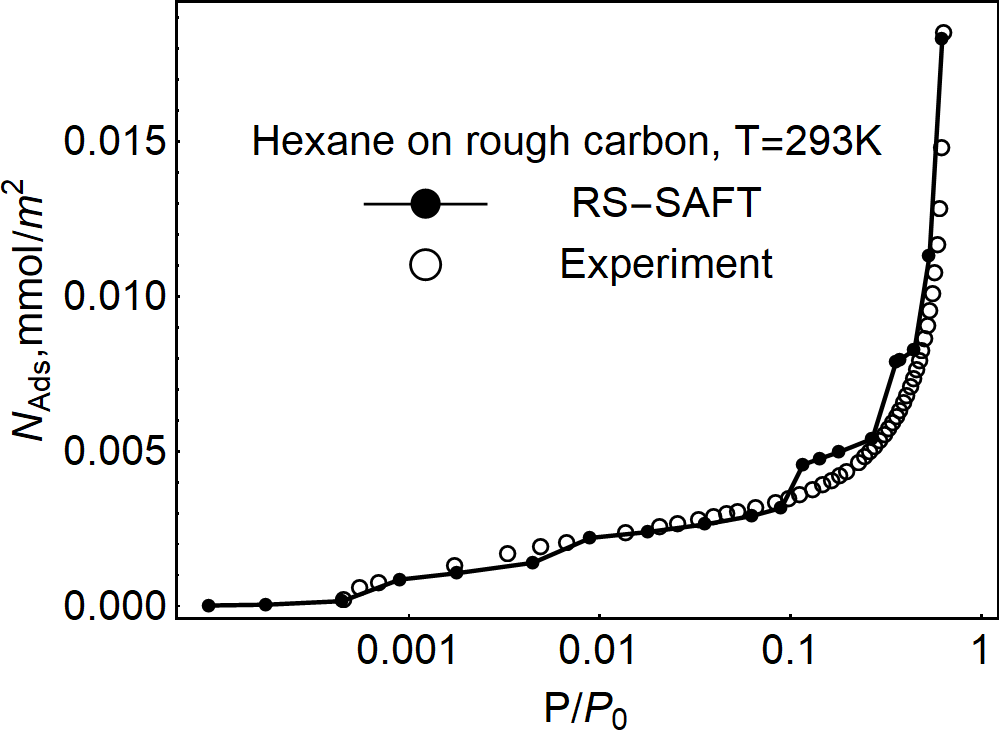}
	
	\includegraphics[width=12cm]{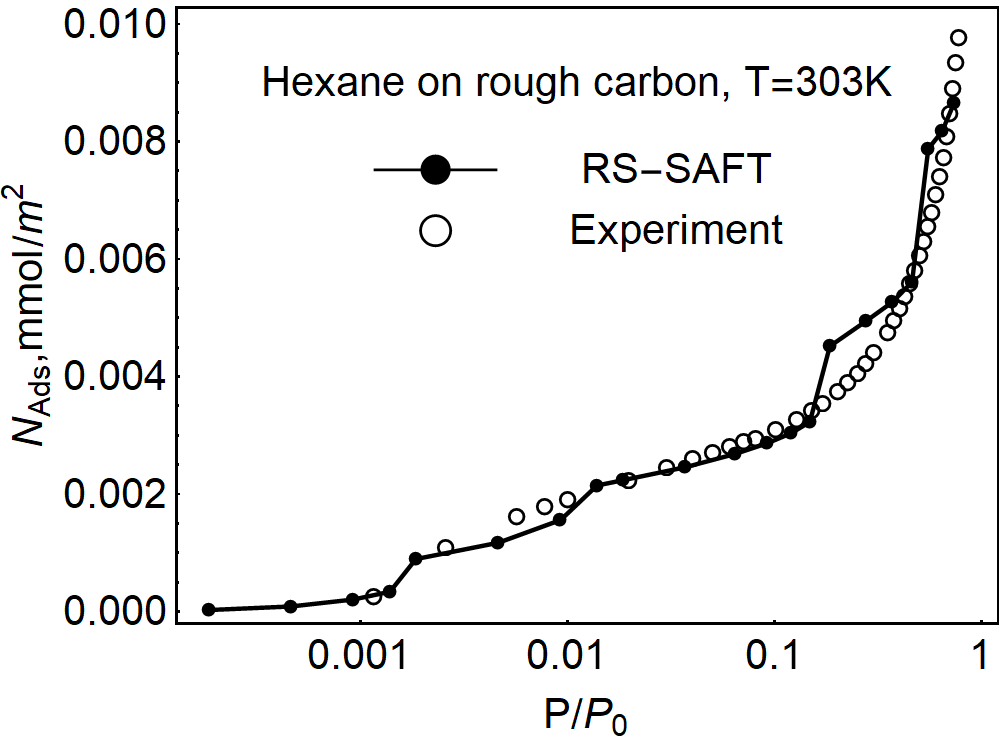}
	\caption{\label{Fig:RS-SAFT} Comparison of adsorption isotherms obtained by experiment and RS-SAFT for hexane on rough carbon at 293K and 303K. Open circles correspond to experimental measurements in our lab for the same solid sample as in the experiment with nitrogen. Solid lines are theoretical results of RS-SAFT obtained for rough surface with  $\delta=0.6$ nm and $\tau=2.1$ nm and parameters from Table.~\ref{Table}}
\end{figure}

We have performed experimental measurements of n-alkane adsorption on the same solid sample as in the case of nitrogen at several temperatures. The high-purity hexane ($>$99.0\%) vapor adsorption isotherms were determined by ASAP2020 using vapor adsorption option. Hexane is placed in a special vapor tube and degassed prior the adsorption measurements.  The set of vapor adsorption isotherms on carbon black surface was measured at 293K and 303K (Fig.~\ref{Fig:RS-SAFT}). The recirculation bath (recirculating chiller with thermostat Julabo FP50) was used to control the temperature of the sample tube along its whole length.

Our developed version of SAFT entitled RS-SAFT allows to use geometrical parameters which are obtained from nitrogen adsorption analysis. Thus, predictions of the theory can be compared with experimental data for hexane. As was noted above the bulk limit of our version of SAFT coincides with very popular SAFT-VR. This fact allows us to use published parameters \cite{garrido2017coarse} for fluid description Table~\ref{Table}. Thus, our version of SAFT can describe bulk PVT properties of hydrocarbons with good accuracy.  
\begin{table}
	\caption{\label{Table} Parameters of hexane for RS-SAFT calculations}
		\begin{tabular}{|c|c|c|c|c|c|c|c|}
			\hline
					&M & $\epsilon_{ff}/k_B$, K&$\sigma$, \AA & $\lambda_r$&  $\lambda_a$ & $\epsilon_{sf}/k_B$, K &$\sigma_{sf}$, \AA \\
			\hline
			Hexane &2 & 376.35 & 4.508 & 6 & 19.26 & 67.2& 3.954 \\
			\hline			
		\end{tabular}
\end{table} 
In order to minimize the number of free parameters we have fixed solid-fluid characteristic diameter in accordance with Lorentz-Berthelot rules \cite{maitland1981} $\sigma_{sf}=\frac{1}{2}(\sigma+\sigma_{c})$, where $\sigma_{c}=3.4 \AA$ is commonly used value of carbon molecule diameter. Geometrical parameters of surface are fixed too and are defined from nitrogen adsorption RSDFT analysis. Thus, we have used only one tuning parameter which is characteristic solid fluid energy $\epsilon_{sf}$.

Comparison of experimental results and RS-SAFT calculation can be found in Fig.~\ref{Fig:RS-SAFT}. As on can see the theory fits experimental data well, especially at small pressures. Also the theoretical predictions correctly reflect temperature changes: the start of adsorption isotherm in terms of relative pressure is significantly shifted to the right with the temperature increase. The most noticeable deviations from experiments the theory demonstrates in vicinity of saturation pressure $P_0$. At these conditions theoretical adsorption isotherms are step-like that is not confirmed by experiments. According to the literature step-like adsorption isotherms relay to well defined structure of adsorbed fluid. Indeed, in our theory all adsorbed chain molecules have parallel orientation to the surface level. On the other hand, pressures near $P_0$ correspond to significant fluid fluctuations which can broke well defined layering. This is exactly what one can see from experimental adsorption isotherms. Despite difference in form of isotherms at pressures near $P_0$ theoretical results have a good quantitative agreement with experiment data.     

\section{Conclusion}
Crossover of DFT approach and SAFT model for chain structured molecules is one of the most perspective way to describe chain fluids surface phenomena. In this publication we have proposed novel version of DFT-SAFT approach entitled RS-SAFT. The key differences of our model are concluded in alternative analytical form for intermolecular term of Helmholtz energy and random surface approach which allows take into account geometrical roughness in both normal and lateral directions.

Developed theory can be applied for prediction of short hydrocarbons adsorption properties on the surface of certain solid sample. We have formulated workflow where information about geometrical roughness is obtained from experiment and RSDFT calculations for low temperature adsorption of simple fluids. As was demonstrated in work \cite{aslyamov2017density} RSDFT allows to construct model of rough surface for certain solid sample which is defined by two geometrical parameters. These surface characteristics are used in further RS-SAFT calculations for n-alkane adsorption on the same solid sample. 

In oder to demonstrate capability of our approach we have compared RS-SAFT predictions with experiment. We have considered hexane adsorption on carbon black sample with initially unknown surface geometry. We have used the same carbon black solid sample in all experiments which conclude: nitrogen adsorption at 77K, hexane adsorption at 293K and 303K. Theoretical RSDFT analysis demonstrates rough geometry of sample.  RS-SAFT predictions for considered surface fit experimental data well. 

\appendix
\section{Hard Sphere RDF}
\label{Appendix}
The initial description of HS fluid was performed by Wertheim \cite{wertheim1963exact} and Thile \cite{thiele1963equation}. They obtained the solution of Ornstein-Zernike  equation in Percus-Yevick (PY) approximation \cite{percus1958analysis}. However, analytical result was obtained only for the Laplace transform $G(t, \phi)$ of product $r g(r, \phi)$ 
\begin{eqnarray}
G(t,\phi)=\int_{0}^{\infty}rg(r,\phi)e^{-t r}dr.
\end{eqnarray}
Wertheim \cite{wertheim1963exact} obtained Laplace image $G(t)$ as analytical function. According to him RDF has the following form:
\begin{align}
\label{RDF_Wertheim}
g(r,\phi)=\frac{1}{2\pi i}	\int_{\delta-i\infty}^{\delta+ i \infty}\frac{t L(t,\phi)e^{t r}dt}{12\phi r \left[L(t, \phi)+S(t, \phi)e^t\right]},
\end{align}
where, $\delta$ is point on the real coordinate of the complex plane, such that $\delta$ is greater than the real part of all singularities of the integrand.
\begin{eqnarray}
\label{LS}
&L(t,\phi)=12 \phi [(1+1/2 \phi) t+(1+2 \phi)], \nonumber 
\\ 
&S(t,\phi)=(1-\phi)^2t^3+6\phi(1-\phi)t^2+18\phi^2t-12\phi(1+2\phi). \nonumber
\end{eqnarray}

Here integral \eqref{RDF_Wertheim} is calculated by direct method using residue theorem of complex analysis. For determination of singularity points it is necessary to solve following equation in respect to variable $t$ (denominator of \eqref{RDF_Wertheim} equals to zero):
\begin{eqnarray}
F(t,\phi)=L(t,\phi)+S(t,\phi)e^{t}=0.
\end{eqnarray}
After substitution of expressions \eqref{LS}, it transforms into transcendental equation for variable $t$:
\begin{eqnarray}
\label{Eqn1}
12 \phi [(1+1/2 \phi) t+(1+2 \phi)]+[(1-\phi)^2t^3+\nonumber \\
+6\phi(1-\phi)t^2+18\phi^2t-12\phi(1+2\phi)]e^{t}=0.
\end{eqnarray}
Equation \eqref{Eqn1} has infinite number of roots on complex plane. Let us start with obvious root $t=0$ which is pole of the third rang, also it is unique real solution of \eqref{Eqn1}. The other roots are conjugated complex simple poles $t_n=R_n\pm i I_n$, where $R_n$, $I_n$ are real and imaginary parts of complex number. Thus, in accordance to residue theorem,  expression \eqref{RDF_Wertheim} can be rewritten as:
\begin{eqnarray}
\label{RDF_Residue}
&g(r,\phi)=1+\dfrac{d}{r}\sum\limits_{\left\lbrace t_n \right\rbrace }\dfrac{t_n L(t_n,\phi)}{12\phi F'(t_n,\phi)}e^{t_n r/d}= \nonumber
\\
&=1+\dfrac{d}{r}\sum\limits_{\left\lbrace t_n \right\rbrace}C_n(\phi)e^{t_n r/d},
\end{eqnarray}
where $ F'(t_n,\phi)$ is derivative with respect to $t$ at the point $t=t_n$, here the first term \textquotedblleft1\textquotedblright corresponds to the residue at point $t=0$, the second term is sum over all simple complex poles. Simpler expression can be obtained after summing conjugated poles:
\begin{eqnarray}
\label{RDF_Sum}
&g(r,\phi)=1+\dfrac{2d}{r}\sum\limits_{n=1}^{\infty}A_n(\phi)e^{R_n r/d}\cos(I_n r/d+\alpha_n), \nonumber \\
\end{eqnarray}
where
\begin{eqnarray}
\label{RDF_A_phi}
&A_n=\left|\dfrac{t_n L(t_n,\phi)}{12\phi F'(t_n,\phi)} \right|,\,\, \,\ \alpha_n=\arg\left(\dfrac{t_n L(t_n,\phi)}{12\phi F'(t_n,\phi)}\right). \nonumber 
\\
\end{eqnarray}
Expression \eqref{RDF_Sum} contains only real functions which are depended on $\phi, t_n$. Thus, as one can see from \eqref{RDF_Sum}, in order to calculate RDF one needs only the distribution of roots $t_n(\phi)$. 

Let us consider new equation which is the limit $|t|\to \infty$ of equation \eqref{Eqn1}:

\begin{eqnarray}
\label{Eqn2}
12 \phi(1+1/2 \phi)+(1-\phi)^2z^2e^{z}=0
\end{eqnarray}
By introduction of a new variable $q=-\frac{12\phi (1+1/2 \phi)}{(1-\phi)^2}$ it is possible to rewrite \eqref{Eqn2} in more simple form: $z^2e^z=q$. Such equation can be solved exactly in terms of Lambert functions $W(x)$ \cite{scott2006general,corless1996lambertw}
$$W(x)e^{W(x)}=x.$$

After simple modifications, the above equation can be written as $(2W(x))^2e^{2 W(x)}=4x^2$. Thus, solution of equation \eqref{Eqn2} has the following form 

\begin{eqnarray}
\label{Lambert_0}
z_n=2 W(n,\pm q^{1/2}/2),
\end{eqnarray}
where $n=1,2,...$ enumerates complex branch of Lambert function. 


Using exact solution \eqref{Lambert_0} as the limit, the solution of \eqref{Eqn1} can be written as series of $z_n^{-1}$:
\begin{eqnarray}
\label{Lambert_Series}
t_n=z_n+\sum_{k=1}^{\infty}a_k z_n^{-k}
\end{eqnarray}
where coefficients $a_n$ depend only on density $\phi$ and can be found after substitution of \eqref{Lambert_Series} in \eqref{Eqn1}. For the aims of this work it will be enough to consider only six terms in the sum \eqref{Lambert_Series}. Corresponding coefficients have the following explicit expressions:

\begin{eqnarray}
& a_1=\frac{2(1-5\phi+5\phi^2)}{(1-\phi)(2+\phi)}, \,\,\, a_2=\frac{2 (-5 + \phi (15 + 2 \phi (15 + 7 \phi)))}{(1-\phi)(2+\phi)^2} \\
& a_3=\frac{4 \left(\phi  \left(\phi  \left(\phi  \left(\phi  \left(26 \phi ^2+222 \phi +309\right)+128\right)+12\right)+6\right)+26\right)}{3 \left(\phi ^2+\phi -2\right)^3} \nonumber \\
& a_4=\frac{2 \left(791 \phi ^8+3020 \phi ^7-358 \phi ^6-9448 \phi ^5-9640 \phi ^4-2506 \phi ^3-550 \phi ^2-886 \phi -106\right)}{3 \left(\phi ^2+\phi -2\right)^4} \nonumber \\
& a_5=\frac{4 \left(-935 \phi ^{10}+4205 \phi ^9+36330 \phi ^8+50790 \phi ^7-15750 \phi ^6-86700 \phi ^5-72180 \phi ^4-23640 \phi ^3-6330 \phi ^2-4000 \phi +112\right)}{5 \left(\phi ^2+\phi -2\right)^5} \nonumber \\
&a_6=\frac{4}{15 (\phi -1)^5 (\phi +2)^6}(-34645 \phi ^{11}-327355 \phi ^{10}-851875 \phi ^9+ 47910 \phi ^8+3331605 \phi ^7+ \nonumber\\ 
&+5742165 \phi ^6+4589340 \phi ^5+ 2083380 \phi ^4+653055 \phi ^3+165055 \phi ^2+19258 \phi -6104) \nonumber
\end{eqnarray}
Thus zeros of \eqref{Eqn1} are defined by the following partition sum:
\begin{eqnarray}
t_n=z_n+\frac{a_1}{z_n}+...+\frac{a_6}{z_n^6}
\end{eqnarray}
The next question is summation over $n$ (index of zeros) in expression \eqref{RDF_Sum}. It is easy to verify, that for all poles $R_n=Re(t_n)<0$, and $R_{n+1}<R_n$, then contribution of exponential n-th term in sum \eqref{RDF_Sum} rapidly decreases, when the number $n$ increases. Thus, for accurate result summation over all terms in \eqref{RDF_Sum} is not required and \eqref{RDF_Sum} can be rewritten as analytical expression with $K$ terms (in our study we have used $K=10$): 
\begin{eqnarray}
\label{RDF_PY_L}
g(r)=1+\frac{2d}{r}\sum_{n=1}^{K}A_n(\phi)e^{R_n r/d}\cos(I_n r/d+\alpha_n).
\end{eqnarray}

Despite wide applications, PY approximation has two weak points: the contact value $g(d)$ is too low at high density; phase of oscillation at large distance $r$ differs from the exact one. In order to correct these artifacts construction of Verlet-Weis can be used \cite{verlet1972equilibrium}. This is achieved by introduction of a modified density $\phi_M=\phi+\phi^2/16$ and modified HS diameter $d_M=\left(\phi_M/\phi\right)^{1/3}d$ in order to correct RDF oscillation. Verlet and Weis, also, proposed an additional term which improved contact value $g(d)$, so corrected RDF has following form:
\begin{eqnarray}
\label{RDF_VW}
&g^{VW}(r/d)=g(r/d_M;\phi_M)+\dfrac{1}{r}e^{\frac{\alpha(r-d)}{d}}\cos\dfrac{\alpha (r-d)}{d}, \nonumber \\
\end{eqnarray}
where parameters $A$ and $\alpha$ can be found from
\begin{eqnarray}
&\dfrac{A}{d}=\dfrac{3}{4}\dfrac{\phi_M(1-0.7117 \phi_M-0.114 \phi_M^2)}{(1-\phi_M)^4} \nonumber \\
&\alpha = \dfrac{24 A/d}{\phi_M g(d_M, \phi_M)} \nonumber
\end{eqnarray}
The form of expression of added term in \eqref{RDF_VW} coincides with analytical result \eqref{RDF_Sum}. This fact helps the process of further calculations of corrected form \eqref{RDF_VW}.
\bibliography{RS-SAFT}
\end{document}